\documentclass[12]{elsarticle}

\usepackage{graphicx,amsmath,amssymb,amsfonts}   
\usepackage{dcolumn}    
\usepackage{bm}         
\usepackage{color}
\usepackage{times,setspace,multirow}

\makeatletter
\long\def\@makecaption#1#2{%
  \vskip\abovecaptionskip
  \sbox\@tempboxa{#1. #2}%
  \ifdim \wd\@tempboxa >\hsize
    #1. #2\par
  \else
    \global \@minipagefalse
    \hb@xt@\hsize{\hfil\box\@tempboxa\hfil}%
  \fi
  \vskip\belowcaptionskip}
\makeatother

\begin{document}
\title{Optimising the neutron environment of Radiation Portal Monitors: a computational optimisation study}

\author[ccfe]{Mark R. Gilbert\corref{cor1}}
\author[ccfe]{Zamir Ghani}
\author[ccfe]{Lee W. Packer}
\address[ccfe]{United Kingdom Atomic Energy Authority, Culham Science Centre, Abingdon, OX14 3DB, UK}

\cortext[cor1]{Corresponding author, mark.gilbert@ccfe.ac.uk}

\begin{abstract}
Efficient and reliable detection of radiological or nuclear threats is a crucial part of national and international efforts to prevent terrorist activities. Radiation Portal Monitors (RPMs), which are deployed worldwide, are intended to interdict smuggled fissile material by detecting emissions of neutrons and gamma rays. However, considering the range and variety of threat sources, vehicular and shielding scenarios, and that only a small signature is present, it is important that the design of the RPMs allows these signatures to be accurately differentiated from the environmental background.
Using Monte-Carlo neutron-transport simulations of a model helium-3 detector system we have conducted a parameter study to identify the optimum combination of detector shielding and collimation that maximises the sensitivity of RPMs. These structures, which could be simply and cost-effectively added to existing RPMs, can improve the detector response by more than a factor of two relative to an unmodified, bare design. Furthermore, optimisation of the air gap surrounding the helium tubes also improves detector efficiency.

\end{abstract}

\maketitle

\section{Introduction}

In recent years, concerns about the threat of nuclear and radiological terrorism and of other malicious acts involving such materials have been raised at the international level~\cite{mindecl}.  Coordinated steps are being taken among states to minimise the threat, but over the next few decades there will be an inevitable global expansion in nuclear technologies and energy. Whilst this will bring significant benefit to society, it will nonetheless result in increased challenges from a nuclear security perspective. A significant number of incidents of trafficking of nuclear materials have been recorded in the IAEA Incident and Trafficking Database (ITDB)~\cite{ITDB}, including material which can be used in radiological devices, and conceivably in nuclear weapons.  For example, a 2013 incident reported by the IAEA~\cite{iaeanewsletter2013} included details of a group of traffickers convicted in Moldova for attempting to sell quantities of uranium and plutonium that had been transported in shielded lead canisters in an attempt to evade detection systems. Furthermore, evidence recently heard in the UK~\cite{LI} refers to the significant illicit transportation of \(^{210}\)Po through European airports and cities, which led to significant contamination of many people and locations. Measures to reduce the radiological and nuclear threat are many-faceted, but an important component includes the ability to directly detect attempts to illicitly transport radiological or nuclear material. Here we focus on passive detection systems that have been employed at the front line, Radiation Portal Monitors (RPMs), which measure the 'threat' response from vehicles as they pass between two detector panels.

Commercially available RPM designs typically use gamma and/or neutron detection technologies to scan vehicles, cargo and people. They have been widely deployed around the world at ports, airports and traffic choke points. Neutron detection systems are, for example, \(^6\)Li-based scintillating glass fibers, \(^{10}\)B-based pressurized gas tubes, or \(^{3}\)He tubes. Similarly, gamma detection systems are based on, for example, polyvinyltoluene (PVT) plastic scintillators,  sodium iodide NaI(Tl) detectors, cadmium-zinc-telluride (CZT) or, at greater expense, high-purity germanium (HPGe) crystals. Here we evaluate neutron detecting RPMs by considering \(^{3}\)He pressurized gas tubes as the detection technology, which are widely used in neutron detectors. They are considered to be a particularly attractive choice because of the high response efficiency to thermal neutrons, due to the high \(^{3}\)He(n,p)\(^{3}\)H reaction cross section, and the very low sensitivity to gamma rays. However, a well known drawback with such systems is the current (lack of) availability of \(^{3}\)He, which is derived from the radioactive decay product of \(^{3}\)H. This is largely due to significantly reduced \(^{3}\)H stockpiles and production sources, which has been accompanied by increased demand~\cite{runkle2011}, and has driven the market to consider alternative neutron detection technologies~\cite{kouzes2015,kouzespnnl,lacy2011}. Nevertheless, given that most, if not all~\cite{runkle2010}, of the neutron-detection systems currently deployed are based on \(^{3}\)He pressurized gas tubes we have used it as the basis of this optimisation study.

Considering the range of radiological or nuclear threat sources, vehicular and shielding scenarios, and that only a small signature is sometimes emitted, it is important that the design of RPMs allows these signatures to be accurately differentiated from the residual background response, caused largely by cosmological radiation -- without an unacceptable number of false or missed alarms. The purpose of this paper is to consider, via a computational modelling approach, certain design factors affecting the performance of \(^3\)He-based RPMs, including the variation in response to the background radiation field, and to suggest simple, cost-effective enhancements to optimise the ability to detect nuclear threats. The aim was to demonstrate that alterations to a generic RPM model design and surrounding environment can enable a significant reduction in the measured neutron background, and hence lower their detection limit to a `threat' neutron source.

Previous works~\cite{tomanin2013,reesczirr2012} have explored the optimisation of \(^3\)He-based detector response within polythene box-like configurations, as used by Kouzes \textit{et al.}~\cite{kouzes2008}, to factors such as neutron spectra, levels of shielding, polythene moderation, tube diameter, and fill pressure. This work, however, uses an alternative generic RPM design, with performance optimisation for a threat source relative to the neutron background environment through shielding and collimation enhancements.

In the following, Monte-Carlo neutron-transport simulations of a model RPM detector system, performed with MCNPX, combined with a variety of `roadway' materials, are used to evaluate the detection response as a function of location environment. Subsequently, a detailed parameter optimisation is applied to simple and cost-effective shielding and collimation modifications to demonstrate that they could be added to existing RPMs to significantly reduce detection limits to a neutron threat-source in an environmental neutron field. Optimisation of the generic RPM design, via the introduction of air gaps, is also considered.

\section{Setup \& Preliminaries}\label{setup}

\begin{figure}
\includegraphics[width=0.5\textwidth]{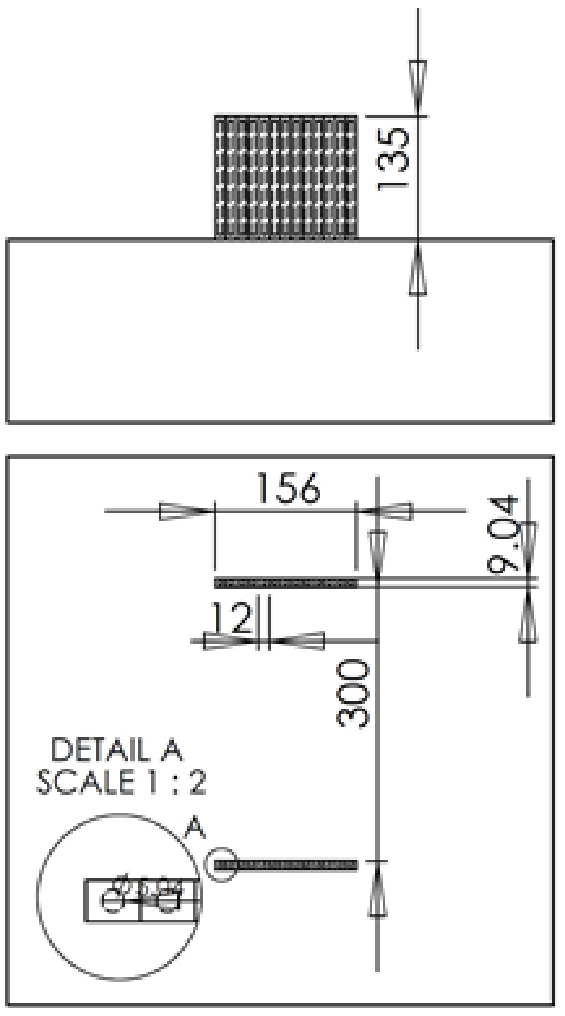}
\includegraphics[width=0.5\textwidth]{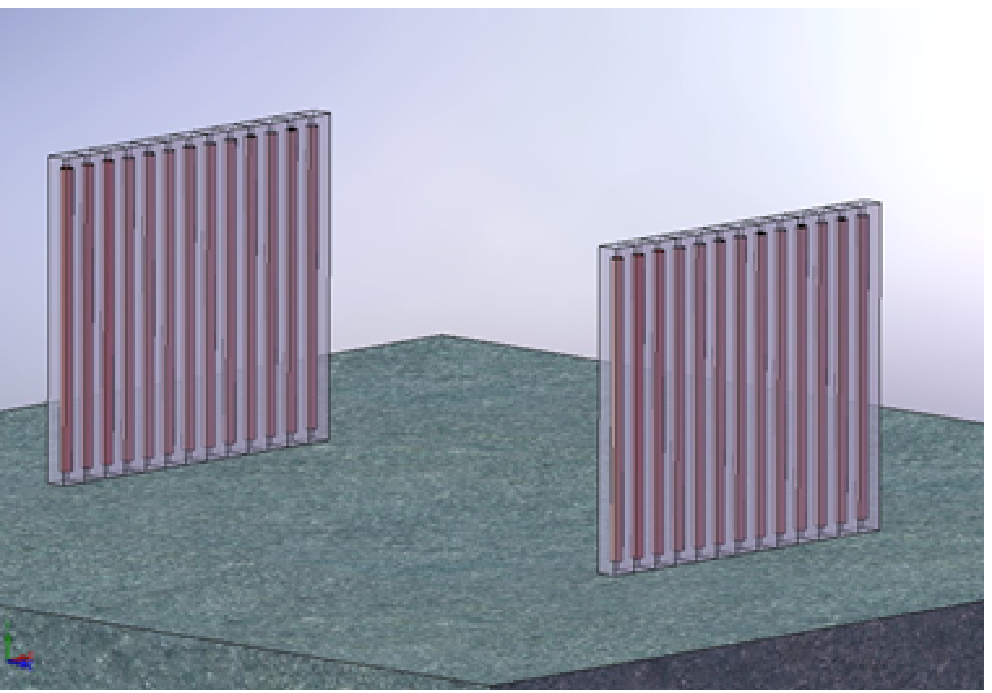}
\caption{\label{detector}(lhs) Generic \(^3\)He-based panel detector model with dimensions (in cm), showing as side view (top) and view from above (bottom). (rhs) 3-D image of detectors.}
\end{figure}

A pair of basic `generic' \(^3\)He-based detector panels were developed for this study. Each RPM panel sits at ground level and consists of an array
of 13 one atmosphere \(^3\)He tubes, of length 125 cm and radius 2.52 cm of which 0.1 cm is the stainless steel tank
enclosing the helium. The \(^3\)He in the model has a density of \(2.5\times10^{-5}\)~atoms~per~barn-cm and the surrounding 304LN stainless steel, whose composition is shown in Table~\ref{groundmaterials},  has a density of 7.9~g~cm\(^{-3}\). The tubes are completely surrounded by polythene (two-thirds H atoms, one-third C) with a density of 0.93~g~cm\(^{-3}\), although later (section~\ref{baseoptimisation} we consider adaptations to this where there is an air gap between the tubes and polythene. Each polythene sub-unit block is of dimension
12 cm x 9.04 cm. The detection panels are 156 cm in length and 135 cm in height. The two panels are spaced by a
distance of 300 cm. The panels are positioned above a ground material volume, which is 100 cm thick, and surrounded by air at a density of \(5.41\times10^{-5}\)~atoms~per~barn-cm.
Fig.~\ref{detector} shows a schematic and 3D view of the basic design, while Fig.~\ref{detectorplaneview} shows more detail, including some of the modifications discussed in later sections.

\subsection{Environmental background neutron sources}

The key performance measure in the present work is the relative response of a given RPM configuration to the defined threat in comparison to the environmental background. Background neutrons are produced in the atmosphere and upper few meters of the earth's crust by interactions between cosmic rays and nuclei. Primary cosmic rays collide with nuclei in the upper atmosphere and create cascades of secondary particles, including neutrons and muons, that can penetrate to the Earth's surface. These high-energy secondary neutrons interact with the local environment at the Earth's surface, generating further neutrons due to spallation events~\cite{runkle2010}.  We have modelled the atmospheric cosmic ray neutron background as a planar source 2.5~m above the ground with a  cos\(^2\)${\theta}$ distribution from the zenith direction. Note that interaction of this background source with the ground material and detector assembly is modelled, and the implications of varying the ground material is explored below. A hemispherical surface source with an isotopic directional distribution was also considered, but no significant change in the results was noted.

The energy spectrum of the background neutrons was derived from EXPACS ver. 2.22~\cite{expacs}. Input parameters were latitude,
51.0809 and longitude 1.1858 with no environmental (ground) parameters. In Fig.~\ref{groundspectra} the `blank' curve shows the EXPACS spectrum implemented.

A background normalisation factor \(N_{bkgrd}\) was calculated for the model using the asphalt pavement ground material
model and by setting a nominal neutron flux of 100 n m\(^2\) s\(^{-1}\)~\cite{kouzes2008} in a region at the
mid-point between the two detectors in this model. This normalisation factor \(N_{bkgrd}\) equals 5950 background neutrons per second, and was applied to all simulations in this study (regardless of the ground material used).

Note that in the present studies there is no consideration of how the background might suppressed due to physical proximity of a vehicle or other structure.

\subsubsection{Background variation with ground material}
The relative neutron background response of an RPM is determined by its immediate environment. In particular, background neutrons will be produced below the RPM from interactions with high atomic mass nuclei in the ground material or `roadway'. Whilst not the primary consideration in the present work, it is interesting to explore how background neutron production can be influenced (minimised) by replacing existing asphalt or concrete based ground surfaces with a material containing reduced concentrations of high atomic mass aggregate materials and increased hydrogenous neutron-absorbing material.

For a variety of different ground materials -- see table~\ref{groundmaterials} -- an MCNPX~\cite{mcnpx} simulation of \(10^7\) particle histories was performed using the environmental source discussed above,  and the \((n,p)\) (neutron absorption followed by proton emission) reaction rate \(R\) was tallied in the \(^3\)He tubes. This reaction, which has a high thermal cross section, produces two charged particles -- a proton and a triton -- whose ionization can be detected and counted via energy deposition pulses~\cite{knoll}. Note that other reactions on \(^3\)He, such as \((n,d)\) (neutron capture followed by deuteron emission), are negligible in the present study, with the \((n,p)\) reaction typically contributing more than 99.7\% of the total reaction rate (also tallied). In MCNPX, which is used for all calculations in the present work, the upper neutron energy was adjusted to \(1\times10^5\)~MeV (\texttt{EMAX} parameter, \texttt{phys:N 1E5}) with the remaining neutron physics options set as default.

\begin{figure}
\includegraphics[width=1.0\textwidth]{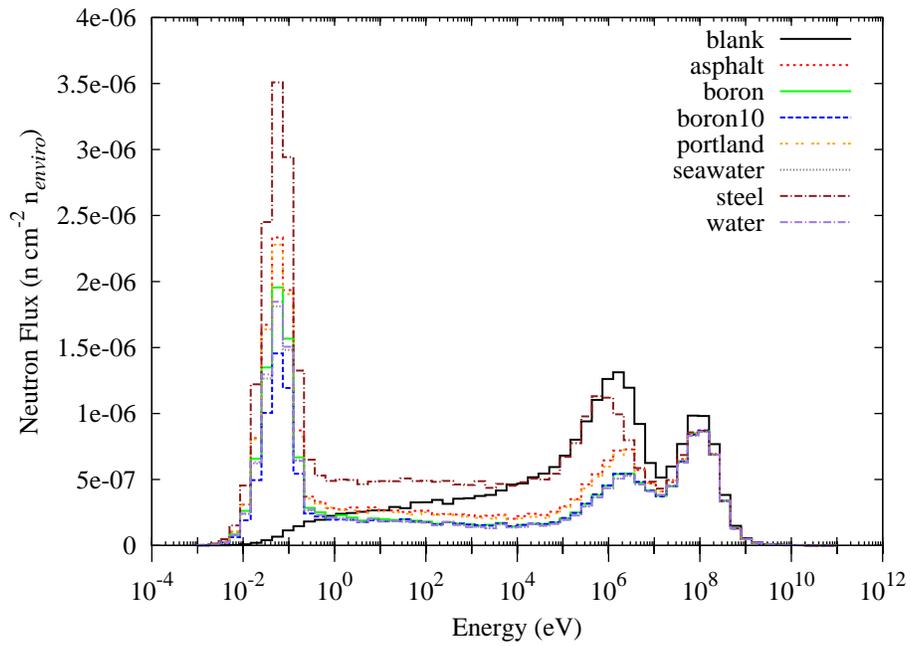}
\caption{\label{groundspectra} Comparison of background neutron spectra experienced in the \(^3\)He tubes with different ground materials (planar
background source). The flux units for each neutron energy are the raw output from MCNPX, which in this case is per environmental neutron n\(_{enviro}\).}
\end{figure}

\begin{figure}
\includegraphics[width=1.0\textwidth]{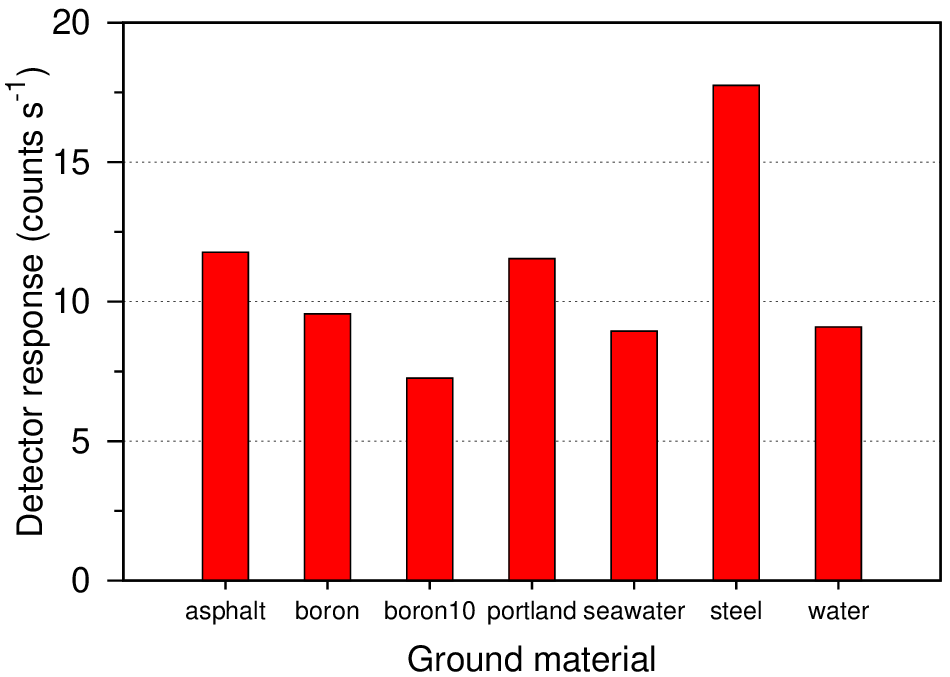}
\caption{\label{backgroundresponse} Comparison of detector responses to the environmental neutron background  for different ground material types.}
\end{figure}

Fig.~\ref{groundspectra} shows the background neutron spectra, measured in the helium tubes of the bare, unshielded detectors, for the different ground materials. The spectra were measured (tallied) in MCNPX using 60 equally-spaced (on a logarithmic scale) energy grid-points between \(1\times10^{-3}\)~eV and \(1\times10^5\)~MeV. The `blank' in the figure is the result in the absence of any ground material. In this `ideal' scenario the proportion of thermal neutrons, which could produce count noise in the detectors, is very low in the tubes. The best performing ground material -- that which keeps the thermal neutron component lowest -- is the \(^{10}\)B-enriched boronated silicone rubber. This is confirmed by the count rate histogram shown in Fig.~\ref{backgroundresponse}, where the enriched rubber produces the lowest background counts per second \(C_{bkgrd}\), calculated by multiplying the background normalisation \(N_{bkgrd}\) (background source neutrons per second) by the MCNPX-calculated \((n,p)\) reaction rate \(R_{bkgrd}\) (reactions~cm\(^{-3}\) per source neutron) and the total volume of \(^3\)He tubing \(V_{He}\) (\(=2296.12\)~cm\(^3\)):
\begin{equation}\label{background_count}
C_{bkgrd}=N_{bkgrd}R_{bkgrd}V_{He}.
\end{equation}

However, as shown in Fig.~\ref{backgroundresponse}, standard asphalt or Portland concrete roadways perform reasonably, certainly compared to steel, and so it is likely that there would not be a great cost benefit to replacing these `standard' roadway materials. This would be especially true if it is desirable for the RPM to be mobile -- far better to improve the shielding of the detector itself to background noise. This is explored further below, with asphalt assumed for the ground material hence forth.

\begin{table}
\caption{\label{groundmaterials} The compositions and densities of the different ground materials considered.}
\begin{tabular}{c|c|p{0.5\textwidth}}\\
Material & Density & Composition (weight \%)
\\\hline
Asphalt Pavement & 2.5784& 45.91\% O, 23.14\% Si, 8.45\% Ca, 7.62\% C, 5.10\% Al + 9.78\% remainder of Fe, Mg, K, Na, H, Ti, S, Pb, N, Mn, V, and Ni \\\hline
304LN Stainless Steel & 7.9 & 70.17\% Fe, 17.23\% Cr, 10.5\% Ni + 2.12\% remainder of Mn, C, N, Si, P, Ti, Cu, Ni, and Mo\\\hline
Portland concrete & 2.3 & 52.9\% O, 33.7\% Si, 4.4\% Ca, 3.4\% Al, + 5.6\% remainder of Na, Fe, H, K, Mg, and C\\\hline
Boronated silicone rubber (BISCO) &1.119 & 47.71\% C, 31.19\% O, 7.87\% H, 4.94\% Na, 4.55\% natural B, 3.56\% N, and 0.18\% Si\\\hline
Boronated silicone rubber (BISCO) -- &\multirow{2}{*}{1.119}& \multirow{2}{*}{as above but with \(^{10}\)B instead of natural B}\\ enriched to 100\% \(^{10}\)B & \\\hline
Water & 1.0 & 11.17\% H, 88.83\% O\\\hline
Sea Water & 1.023343 & 11.17\% H, 88.83\% O, 1.94\% Cl, 1.08\% Na + 0.31\% remainder of Mg, S, Ca, K, and F
\end{tabular}

\end{table}

\section{RPM design optimisation}\label{optimisation_study}

The above analysis illustrates how sensitive the response of a RPM system can be to the local environment, which can also include the other structures that make up a detector panel.

The approach taken in the following was to consider making simple and cost-effective changes (additions) to the bare, generic RPM design introduced earlier and assess how the particular geometrical parameters of these modifications might influence the response of the detector to both a threat source and background neutron field. In particular, the relative change in response to the two different neutron sources is significant -- if a modification dramatically reduces both counts, then the sensitivity of the RPM to a threat may not be improved. Only in situations where the background count is reduced to a relatively greater extent than the threat count are the detection limits and sensitivity improved. The particular relationships between the two responses, which necessitates at least two MCNPX simulations for each configuration, will be discussed later.

\begin{figure}
\includegraphics[width=1.0\textwidth]{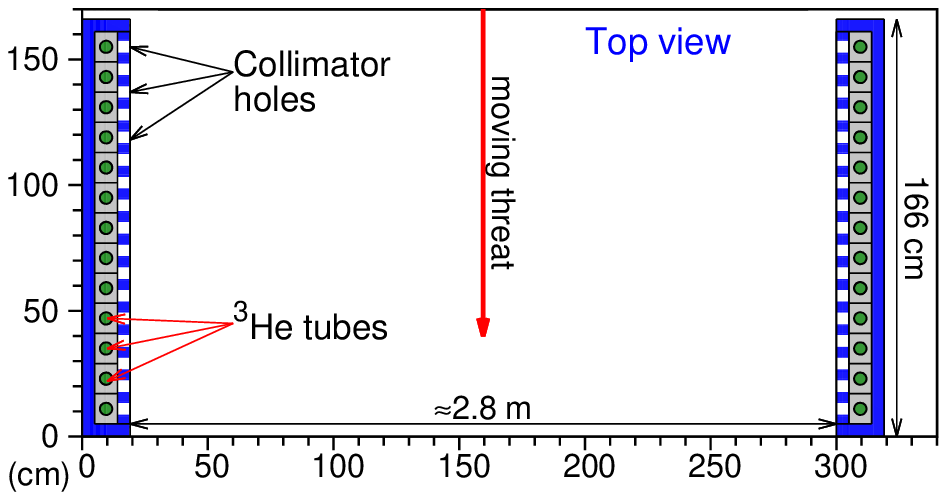}
\includegraphics[width=1.0\textwidth]{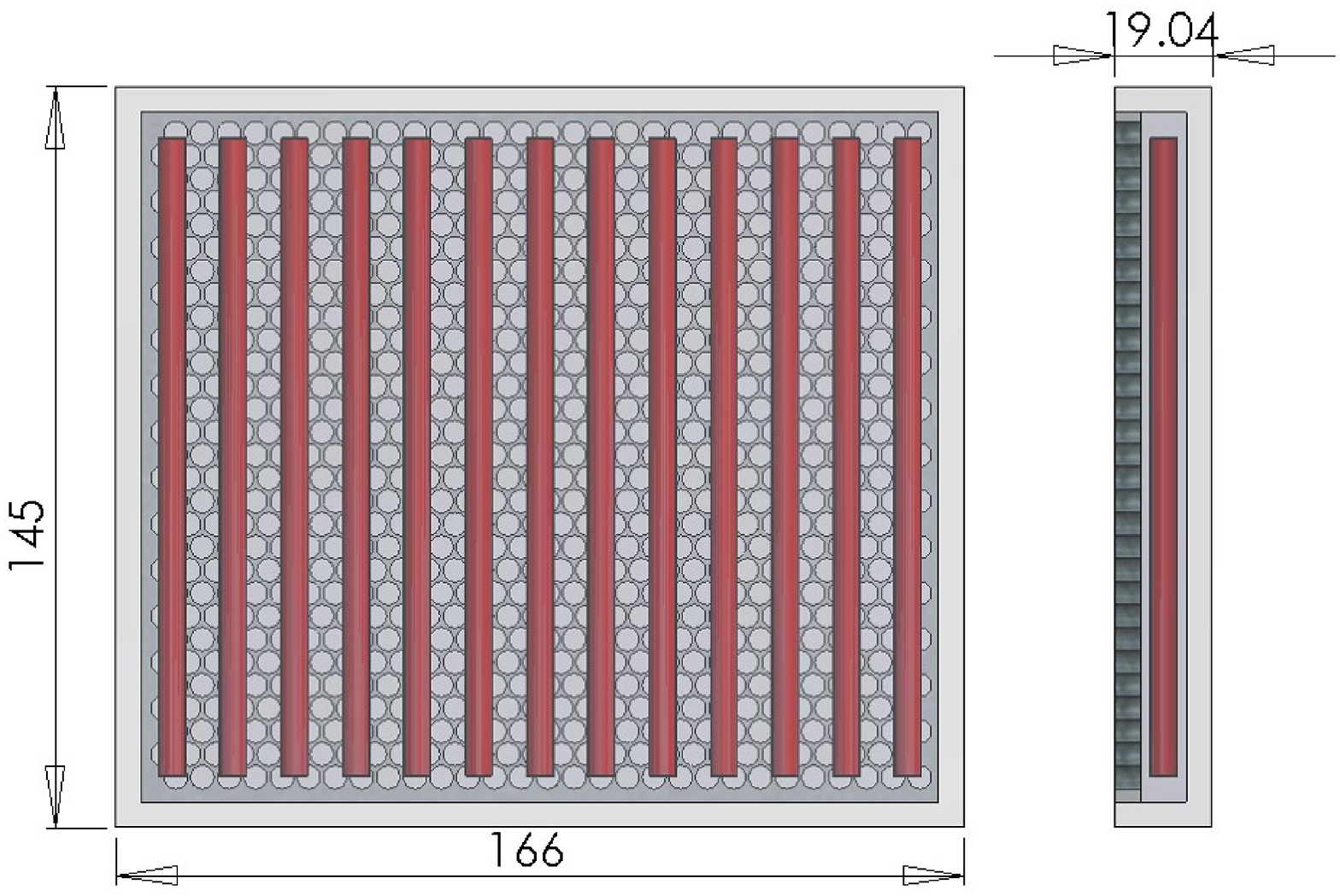}
\caption{\label{detectorplaneview}Top: Diagram (from above) of the model used in MCNPX showing the 2 detector panels with 13 \(^3\)He tubes (green regions)
surrounded by polythene (grey) and the basic configuration of possible boronated polythene shielding and collimator (blue). Note that the addition of the collimator in front of the detectors reduces their overall separation (to 2.8~m here) compared to the original 3~m.  Bottom: 3D renders of an individual detector panel, with the helium tubes shown in red.}
\end{figure}

Several different options for modifications are possible, but for the present study we consider block shielding of varying thickness around the sides and back of the detector. This is combined with a similar shielding block on the front face (facing the `threat') with holes of varying depths and shape, thereby creating a collimation effect, which favours the detection of neutrons approaching from near-perpendicular directions (i.e. from the threat). The base design, with a default 5~cm thick shield and collimator layer of boronated polythene (10\% by mass of boron with a density of 1.0~g~cm\(^{-3}\)~\cite{mcconn}), is illustrated in Fig.~\ref{detectorplaneview}, which is a graphical realisation of the MCNPX model used in the study.

As a first step, the response of a 5~cm thick shielded configuration, with and without a similar 5~cm front-face collimator, was tested against the original bare generic detector design. For this case the collimator holes  are
cylindrical with diameter 4.6 cm spaced approximately 5.2 cm apart (centre-to-centre). MCNPX simulations were performed, firstly to obtain the statistical response of the configuration to the background source, and then secondly, to predict the response of the detector to a \(^{252}\)Cf neutron point `threat' source with an emission rate of 2000~n/s moving at a typical vehicle velocity of 2~m~s\(^{-1}\) (rounded to 1 sig. fig. from that reported in~\cite{ansi4235}) along the mid-line between the two detectors. In the latter case a series of runs were performed with the threat in various discrete locations along the trajectory, the results from which were then integrated in time assuming a 2~s acquisition time. Previous works~\cite{schroettner2009,wahl2007} have considered the impact of varying the vehicle velocity and integration time, but this is beyond the scope of the present work. In all runs \(10^7\) particles histories were calculated, which resulted in reaction rates with statistical errors of less than 3\%. Note that the symmetry of the problem was utilised so that for a threat moving between \(\pm2\)~m relative to the half-way point along the length of the detectors in the 2~s, only 1~s of the acquisition in the positive measurement positions needs to be considered with post-multiplication by 2 to obtain the total response.

\begin{figure}
\includegraphics[width=0.49\textwidth]{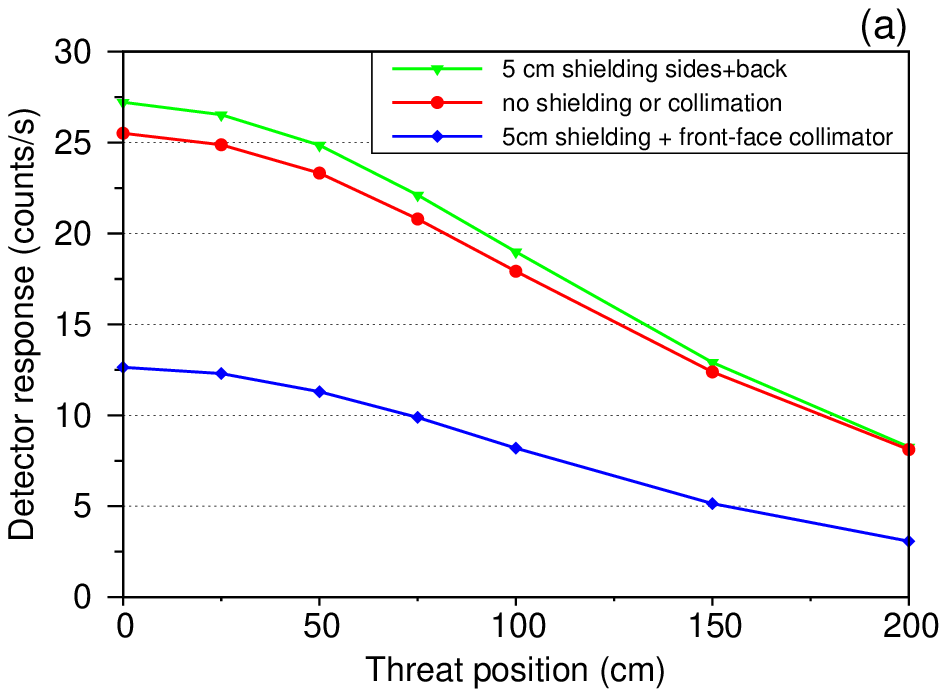}
\includegraphics[width=0.49\textwidth]{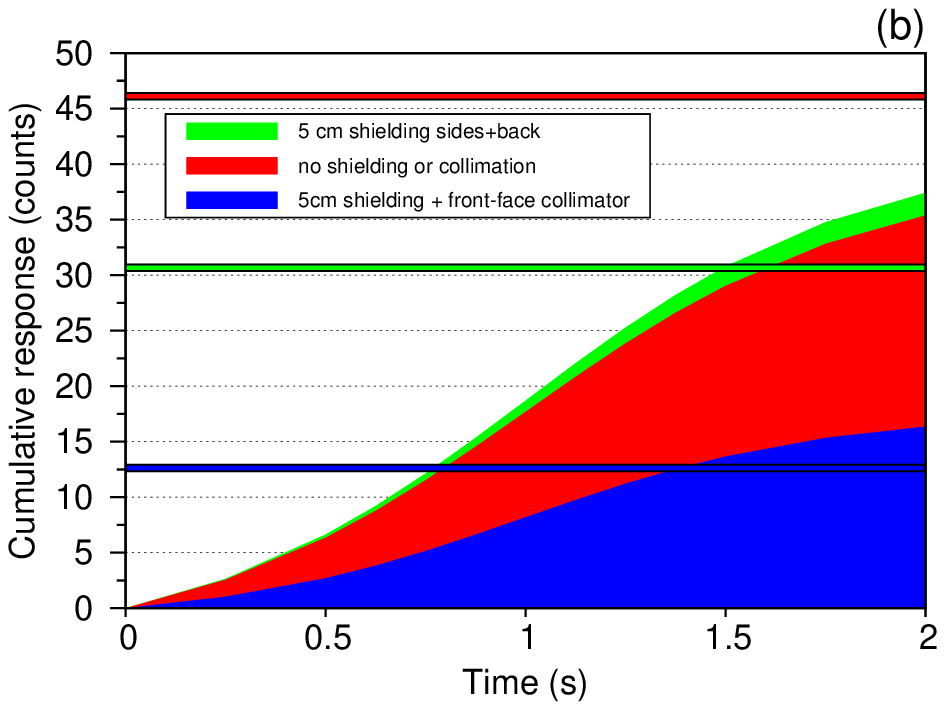}
\caption{\label{response} Source responses for the unshielded RPM, and the default, 5~cm thick shielded and
shield+collimator configurations. The assumed ground material was asphalt. In (a) the absolute response in counts/s of the system as a function of position of the 2000 n/s \(^{252}\)Cf source relative to the half-way point along the length of the detector (position=0), and (b) the cumulative response over the 2~s acquisition time from -200~cm to +200~cm. The horizontal lines in (b) in the same colour as the cumulative fill show the associated alarm level for each of the cases, which is defined by the relative response to the background.}
\end{figure}

Fig.~\ref{response} depicts the results from this initial `test' simulation. In Fig.~\ref{response}a, the response of the three configurations at each of the discrete positions is shown in units of counts per second, calculated via:
\begin{equation}
C=\phi_{src}R_{src}V_{He},
\end{equation}
where \(\phi_{src}\) is the threat source emission strength in n/s, \(R_{src}\) is the \((n,p)\) reaction
rate measured by MCNPX in the helium tubes (reactions per cm\(^3\) per neutron).
As expected, the RPM
response decreases as the position increases, as the source gets further away from the detector. The figure further shows that the shielded configuration increases the count rate compared to the original detector, which is due to increased back-scatter and moderation of the source neutrons. With collimation, on the other hand, the count response is dramatically reduced, suggesting that many of the neutrons are prevented from reaching the detector tubes by the boronated polythene in the collimator. However, this is not necessarily an indicator of a poor RPM design because, as stated previously, it is the relative change in response compared to the change in response to the background that determines detector efficiency.

Fig.~\ref{response}b considers this relative response by comparing the cumulative response of each RPM design over the 2~s acquisition of a threat moving uniformly between -2~m to +2~m to an alarm-level \(A\) set using the Currie limit:
\begin{equation}\label{currie_limit}
A=C^{total}_{bkgrd}+4.65\sigma_{bkgrd},
\end{equation}
where \(\sigma_{bkgrd}\) is the standard deviation of the total background count \(C^{total}_{bkgrd}\) in 2s, which is defined by multiplying the left-hand side of Eq.~\eqref{background_count} by \(t=2\)~s.

From the figure, we can clearly see that, for these set of three default designs with a threat strength of 2000 n/s, the unshielded design
would not detect the threat, while the shielded designs, with and without collimation would, despite the fact that one has a higher  and one a much lower threat response compared to the bare design. This confirms that a sensitive trade-off exists between: 1) achieving a low detector response to the
environmental neutron field and 2) minimising the loss of response to a `threat' source. The following parameter studies investigate this trade-off in more detail.

\subsection{Parameter study}\label{parameterstudy}
The above confirms that the addition of shielding and collimator
can improve detector performance, but to find an optimum configuration a parameter study is needed. In the following, shielding and collimator thickness, as well as collimator hole (`packing') density and profile have been varied and the
resulting detector response compared against an overall system performance function, `minimum source to alarm' (or
`minimum source alarmed' -- see below).

\subsubsection{Performance measure: `minimum source to alarm'}

In a parameter study, a relatively large number of
separate responses must be assessed -- one for each parameter set -- which makes a detailed, counts-vs-time or
response-with-position analysis impractical. Thus, for this extended study we have used a single performance measure
that can be used to compare different detector configurations.

Firstly, rather than considering the variation in response with the source position (as in Fig.~\ref{response}a) we only measure the
overall response to a line source representing all possible source positions during the 2s acquisition time. Not only
does this reduce the number and complexity of the required Monte-Carlo simulations, but it also means that the threat
source response for a given configuration is a well-defined function of a single quantity extracted directly from the
MCNPX simulations. The total source counts per acquisition \(C^{total}\), is given by:
\begin{equation}
C^{total}=\phi_{src}R_{src}tV_{He},
\end{equation}
where  \(t\) is the total acquisition time (2s).

A second Monte-Carlo simulation gives the response of the detector to the planar background, again as an \((n,p)\)
reaction rate \(R_{bkgrd}\), which is used to define the alarm-level \(A\) via the Currie~\cite{currie1968} detection limit defined earlier (see Eq.~\eqref{currie_limit}).

To define the relative
effectiveness of a particular detector configuration, we then define the `minimum source to alarm' as the lowest
theoretically detectable threat strength \(\phi^{min}_{src}\) by equating \(C^{total}=A\), using the square root of \(C^{total}_{bkgrd}\) for \(\sigma_{bkgrd}\), and rearranging. For comparison,
note that the values of this performance measure for the configurations considered in the preliminary calculations in the previous section are 2638 n/s for the bare, unshielded system, 1647 n/s for the system with 5 cm of boronated
polythene behind and at the sides of the detector (which remains fixed for the parameter study in this section), and 1286
n/s for the shielded detector with the 5 cm thick collimator made of boronated polythene.

\subsubsection{Variable parameters}\label{parametervariables}

\begin{figure}
\includegraphics[width=1.0\textwidth]{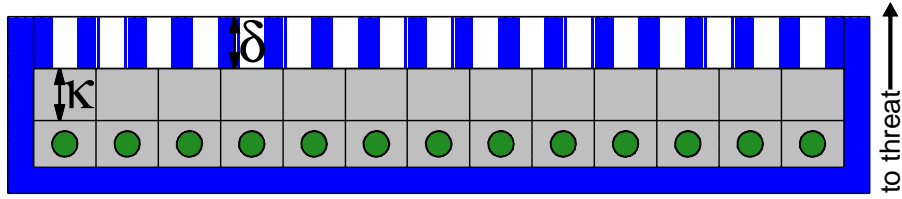}
\caption{\label{top_view_delta_kappa}Top view slice of the single detector configuration used in optimisation study, showing
variable parameters $\delta$ (collimator depth) and $\kappa$ (front shielding thickness). In this picture the threat
neutrons come from above.}
\end{figure}

\begin{figure}
\includegraphics[width=0.7\textwidth]{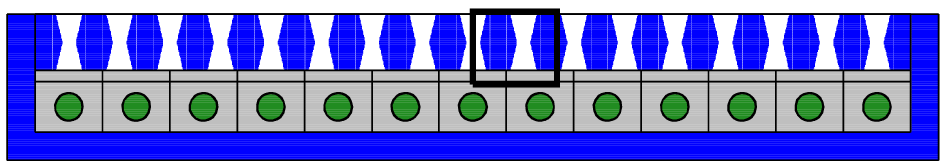}
\includegraphics[width=0.25\textwidth]{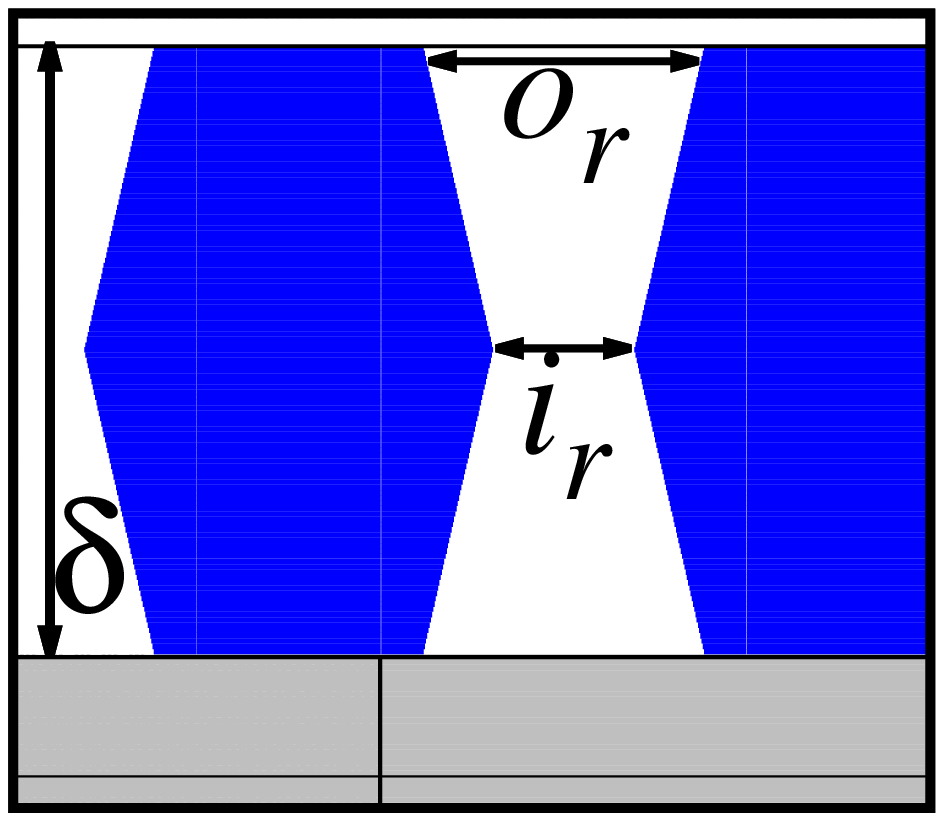}
\caption{\label{top_view_delta_profile}Top view slice of the detector showing optimisation parameters for collimator profile.
}
\end{figure}

Schematics of the detector geometry used for this optimisation study are shown in Fig.~\ref{top_view_delta_kappa}  and Fig.~\ref{top_view_delta_profile}. Note that here we
have taken advantage of further problem symmetry and have only modelled a single detector, locating a reflecting plane on
the other side of the source line distribution (see previous section). This simplification not only makes the input
file for the MCNP model easier to manipulate and adjust for different parameter sets, but it also reduces computational
time, which, as already noted, is crucial when a large number of simulations are required.

In Fig.~\ref{top_view_delta_kappa}  the basic detector design is shown and two of the variable parameters are labelled. The first of these is the
collimator depth $\delta$, which varies from 0.1 to 40 cm. Additionally, an extra layer of solid forward shielding, with thickness $\kappa$, has been added between the detector tubes and collimator, which may improve neutron moderation and detection. $\kappa$ is allowed to range from 0 to 10 cm. Note that the
material used for this extra shielding is the same basic polythene (see section~\ref{setup}) that surrounds the helium tubes, and is not the
boronated polythene used for the collimator and other shielding.

Two other parameters were varied. The first of these, illustrated by Fig.~\ref{top_view_delta_profile}, is the profile of the
holes in the collimator. By default these are simple cylindrical holes, which could be easily drilled
into a sheet of the boronated polythene used to construct the collimator. For the parameter study, the cross sectional
profile of the holes was varied using the ratio
\(i_r/o_r\), where \(i_r\)
and \(o_r\) are the inner (middle) and outer radii of the hole, respectively. This allowed
the profile to vary from cylindrical (\(i_r/o_r=1\)), to
progressively more conical profiles (as illustrated in the figure), where
\(i_r/o_r\) is less than one.

The final parameter was the packing density of the collimator holes, varied by changing
\(o_r\), and measured as the percentage open area on the front face of collimator.

\subsection{Optimisation Results}\label{optimisation_results}

\begin{figure}
\includegraphics[width=1.0\textwidth]{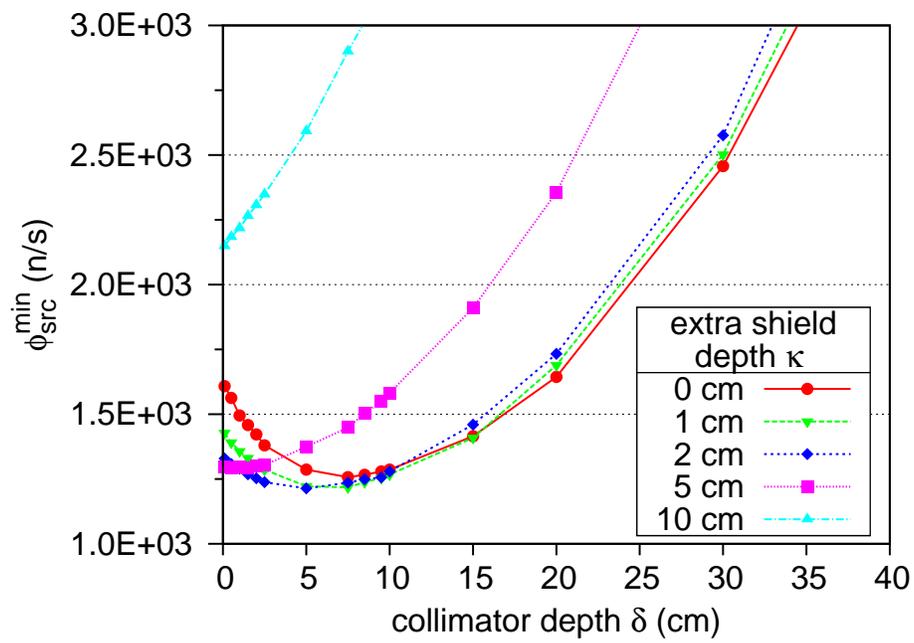}
\caption{\label{study_deltavskappa}Variation in minimum source to alarm as a function of collimator depth $\delta $ for
different thicknesses $\kappa $ of extra forward shielding.}
\end{figure}

In the present we began by
optimising the additional forward shielding thickness $\kappa$, in conjunction with collimator depth $\delta$. From Fig.~\ref{study_deltavskappa},
which shows how the minimum detectable threat \(\phi^{min}_{src}\) varies with $\delta$ for various choices of $\kappa$, we see that there
is an optimum at 1-2~cm for this extra shielding, under the assumption that the collimator depth is close to optimum.
Note that, for very thick collimators, sufficient attenuation is achieved without the need for extra shielding, and so
\(\kappa=0\) is optimal (red curve with circles in figure). However, very large $\delta$ are not likely as these do not produce
good overall detector performance.

With $\kappa=1$ or 2~cm the results demonstrate that a collimator thickness of 5 cm is optimal
producing a minimum alarmed source of approximately \(1.2\times10^3\) n/s, with $\kappa=2$~cm marginally better at this
collimator depth. For this reason $\kappa=2$~cm was selected as the fixed forward shielding thickness in the remainder of this
parameter study, where $\delta$, \(o_r\) and \(i_r\) are
allowed to vary. This leads to a total of 4 cm of polythene between the edge of the stainless steel tanks enclosing the helium and
the start of the collimator, which is an important number to note for the analysis later of the optimum base detector design (section~\ref{baseoptimisation}).

\begin{figure}
\includegraphics[width=1.0\textwidth]{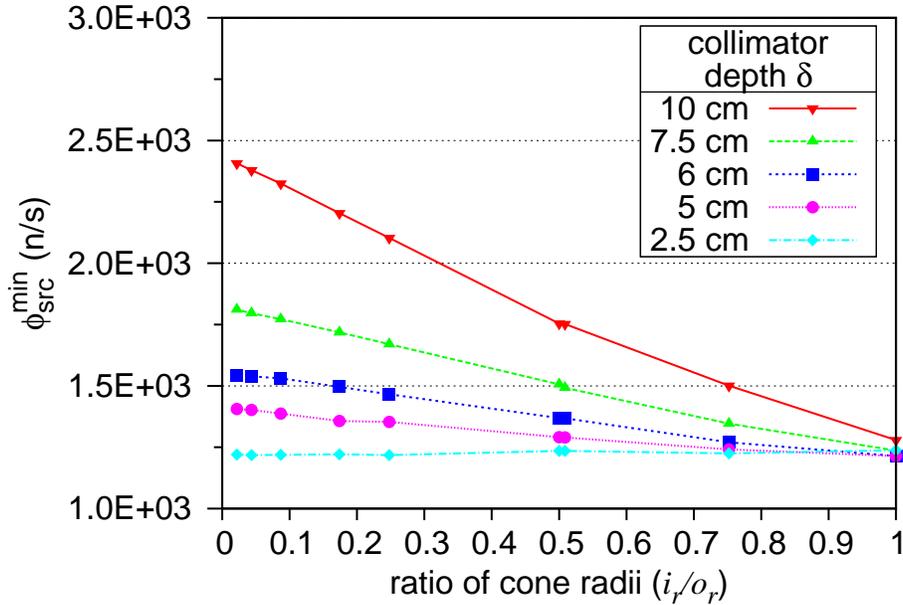}
\caption{\label{study_ratiovsdelta}Variation in detector response as a function of the ratio of the inner or middle radius
of the collimator holes to the outer radius. Curves are shown for different values of the collimator depth parameter $\delta$.}
\end{figure}

Fig.~\ref{study_ratiovsdelta} shows the variation in minimum alarmed source \(\phi^{min}_{src}\) as a function of the ratio \(o_r/i_r\) for different values of $\delta$,
demonstrating that, almost without exception, keeping the holes in the collimator cylindrical (\(o_r/i_r=1\)) is the best option. Note
here that we are only considering a small subset of the collimator depths evaluated earlier because, as Fig.~\ref{study_deltavskappa} showed,
the best detector performances are obtained when $\delta$ is less than around 10 cm.

For the very thinnest collimator configurations ($\delta<2.5$~cm), there may be a very small benefit to having a
more conical profile to the holes, but the minimum alarmed source remains \(\sim1.2\times10^3\)~n/s in all cases. Since thin collimators have relatively little polythene to help with the attenuation of the neutrons,
it is not surprising that any measure which increases the amount of material, such as tapering the holes (\(i_r<o_r\), could produce
improved detector performance. However, a thicker collimator, such as the default 5 cm (red curve with circles in the figure), can produce similar performance with cylindrical holes, so it will be the cost-benefit balance
that determines the best overall preference -- i.e. is less boronated polythene with a complex lattice of gaps cheaper
to make than using more polythene with simple drilled holes?

\begin{figure}
 \includegraphics[width=1.0\textwidth]{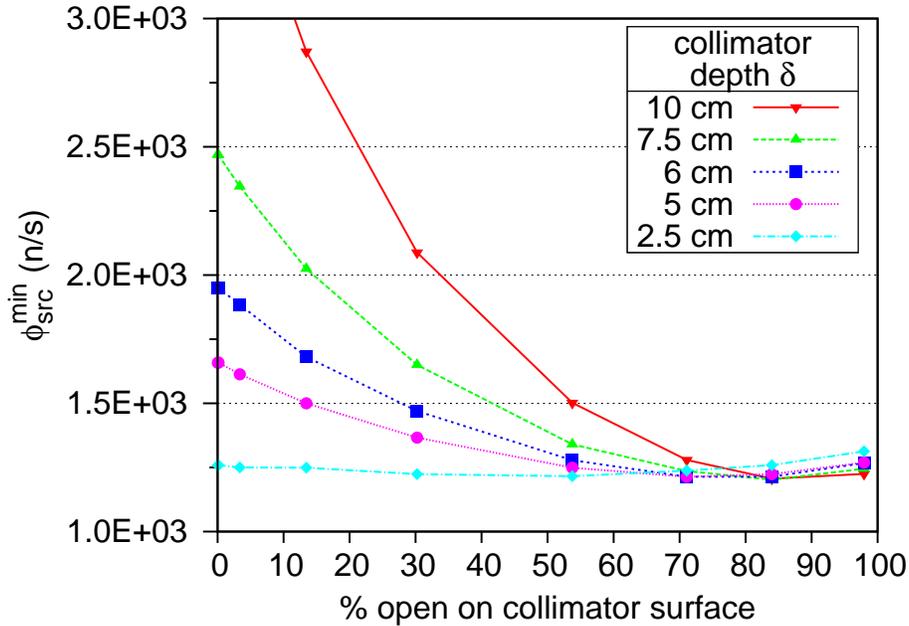}
\caption{\label{study_packingvsdelta}Variation in detector performance as a function of the packing density of the collimator
holes, measured as the percentage of (open) space occupied by the holes on the exposed face of the collimator, with
boronated polythene making up the remainder. Curves are shown for various values of the collimator depth $\delta$.
}
\end{figure}

The final optimisation parameter considered was the packing density of the holes in the collimator themselves. Fig.~\ref{study_packingvsdelta}
shows the detector performance results as a function of the percentage of the front surface of the collimator (facing
the threat) occupied by the holes, which, in the MCNP model, is controlled by varying
\(o_r\). In this case, the value of \(i_r\) is kept
equivalent to \(o_r\), to maintain the optimum cylindrical hole profile identified above. For all curves in
the figure, which represent different collimator depths, the lowest achievable alarmed source is once again \(1.2\times10^3\)~n/s, which has been remarkably consistent throughout the parameter study. This possibly indicates
that this is the lowest achievable result for the current choice of detector design -- i.e. a fixed number of helium
tubes with constant spacing -- and threat type.

Fig.~\ref{study_packingvsdelta} shows that having large cylindrical holes, producing around a 70-80\% total open surface area (the default design considered previously with \(o_r=4.6\)~cm had 71\% packing), is the optimum configuration, except for very thin collimators (the \(\delta=2.5\)~cm results in the figure) where increasing the amount of moderating material through smaller holes can improve matters. While the improvement is only small for this thin collimator configuration, having smaller holes in the thicker collimator designs is significantly undesirable.

\subsection{Optimisation summary}
In summary, for the parameter study, an extra 2 cm thick polythene
shield in front of the basic detector design improved the performance of the RPM, and that additionally a collimator of depth 5 cm with
cylindrical holes occupying about 70-80\% of the exposed surface leads to a minimum source to alarm of \(1.2\times10^3\)~n/s. Below, we consider if this consistent `minimum' can be improved by changing the configuration of the base helium-tube panels themselves.

\section{Base He-3 detector optimisation}\label{baseoptimisation}

It is known that an `albedo cavity'~\cite{tomanin2013} increases the efficiency of a moderated \(^3\)He detector system by promoting additional scattering and reflection of neutrons in the vicinity of the gas tubes. To optimise this effect a study has been carried out on an air gap and moderator configuration for a single detector unit to identify the parameters that produce the greatest gain in signal response relative to the baseline model used above, which contained no air cavity. No background shielding or ground materials were modelled in the following.

\begin{figure}
 \includegraphics[height=0.7\textwidth]{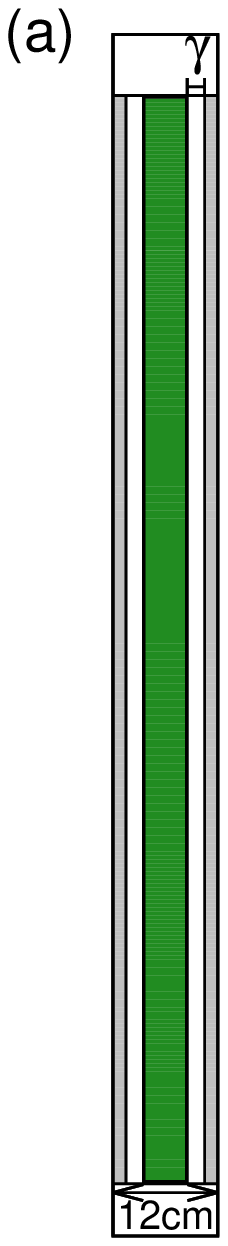}
\includegraphics[height=0.7\textwidth]{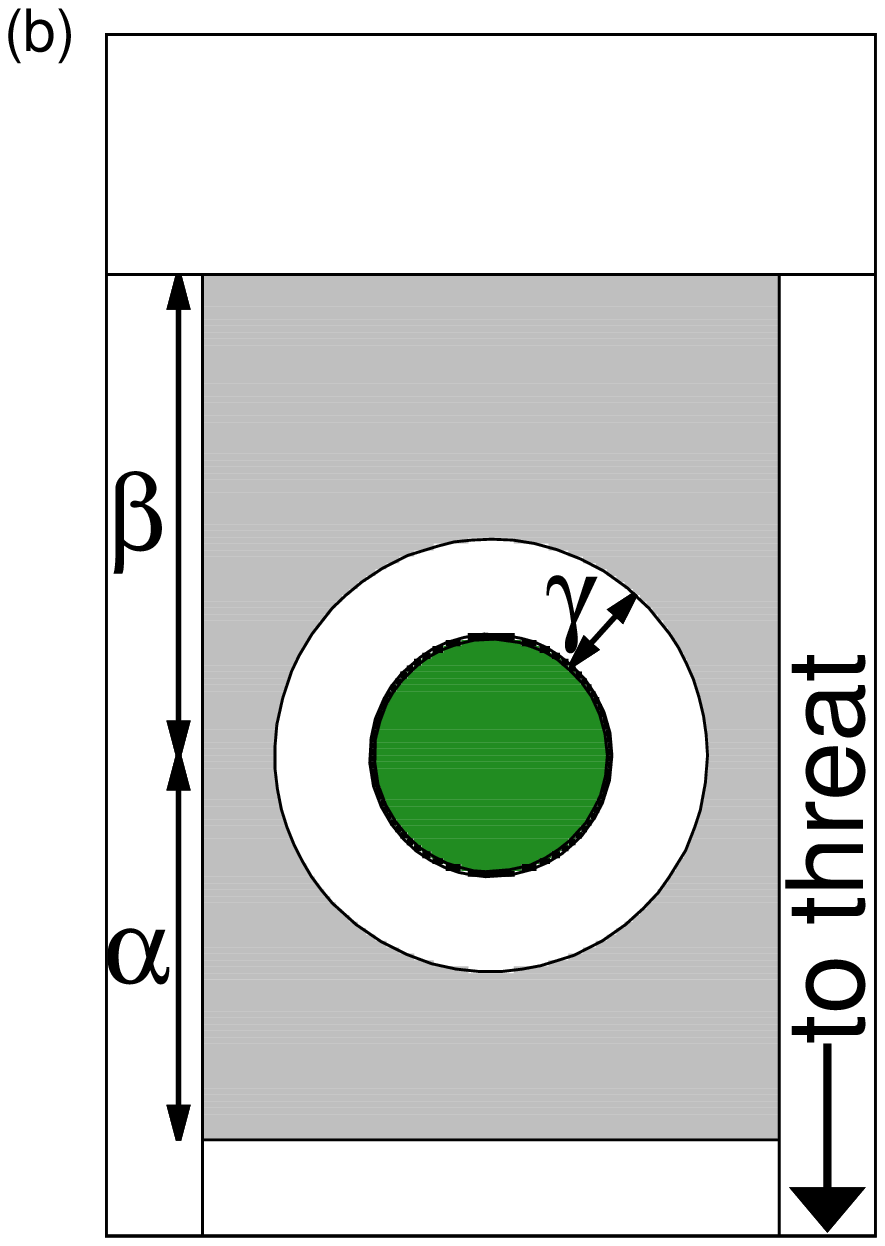}
\caption{\label{cylindrical_gap}(a) Vertical (parallel to detector face) and (b) horizontal  cross sections through the MCNP geometry,
showing a simulated cylindrical shell air cavity between the helium tube (green) and polythene shielding (grey).
}
\end{figure}

\begin{figure}
 \includegraphics[height=0.7\textwidth]{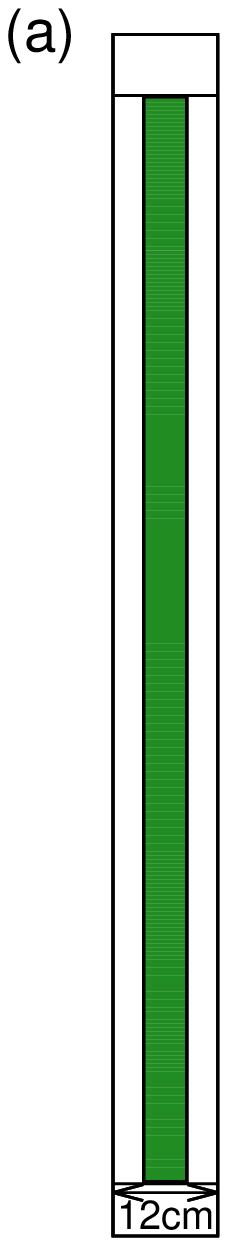}
\includegraphics[height=0.7\textwidth]{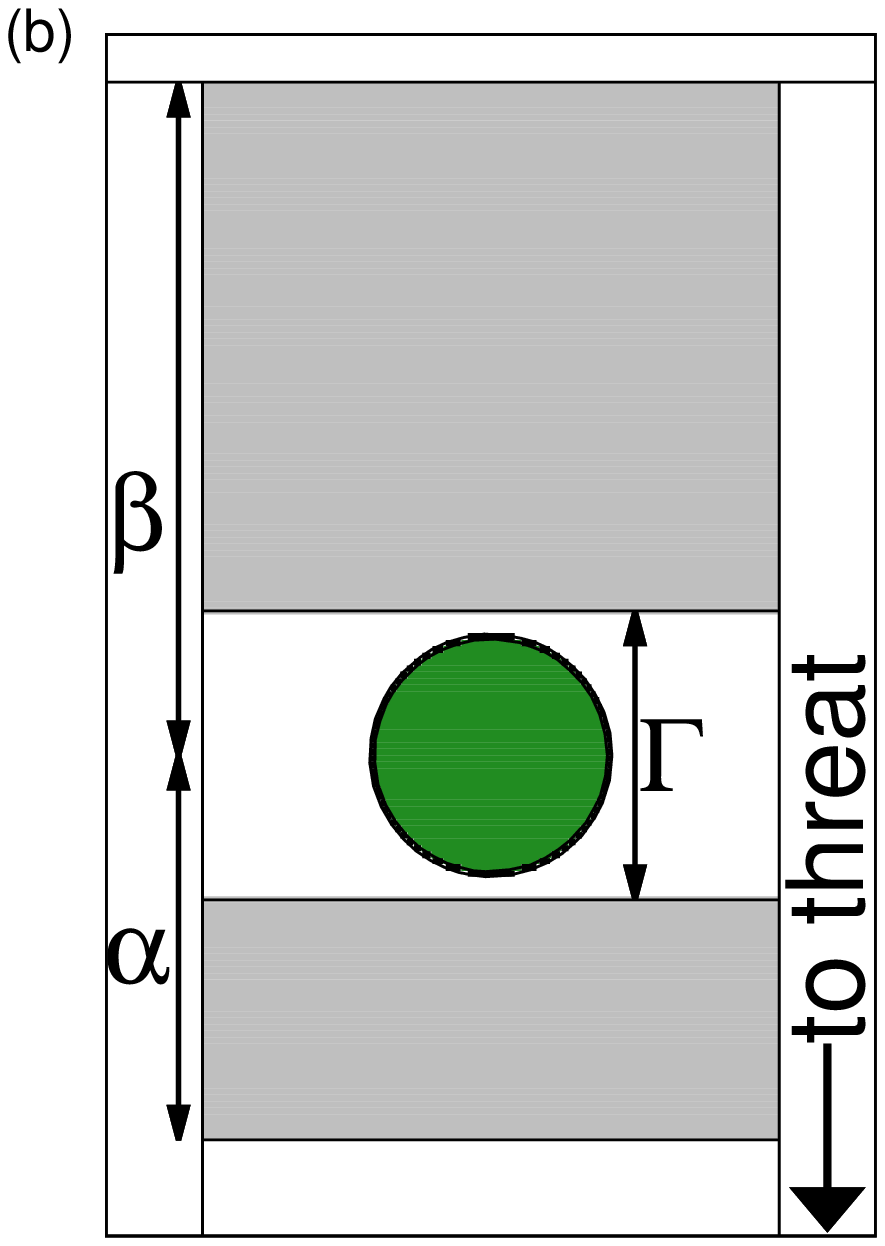}
\caption{\label{square_gap}(a) Vertical (parallel to detector face) and (b) horizontal  cross sections through the MCNP geometry,
showing a simulated rectangular cuboid air cavity containing the helium tube (green) and surrounded by polythene shielding (grey).}
\end{figure}

Two distinct detector geometries have been investigated:
\begin{itemize}
\item The detector volume surrounded by an air cavity consisting of a uniform cylindrical shell, see Fig.~\ref{cylindrical_gap}.
\item The detector volume encompassed by a rectangular cuboid air cavity, which is uniform around the detector volume, see
Fig.~\ref{square_gap}.
\end{itemize}

Air cavity optimisation simulations were carried out such that only a single detector element was modelled with
reflecting boundaries, fixed at 6 cm either side of the centre of the detector. This allows for the simplified
simulation of a bank of detectors with each detector element having a fixed width of 12~cm.

The mcnp\_pstudy software tool~\cite{brown2004} was used to automate the optimisation of the shielding and cavity design study using
MCNPX~\cite{mcnpx}. The tool was used to create input decks by varying the geometric parameters defining the cavity size (width), the depth \(\alpha\)
of front polythene shielding between source and detector and the depth \(\beta\) of rear polythene shielding on the far side of the detector relative to the
source.
For the cylindrical shell cavity the cavity width \(\gamma\) (see Fig.~\ref{cylindrical_gap}) ranged from 0 to 3.5~cm, while for the rectangular gap geometry the total width \(\Gamma\) ranged from 5.04~cm (the diameter of the helium tubes) to 12~cm. Note that the cavity always extends from the base of the cylindrical detector to its top. For the cylindrical cavity set-up the front \(\alpha\) and rear \(\beta\) shielding depths, both measured relative to the centre of the detector, ranged from 2.52 (helium tube radius) to 10~cm. For the rectangular set-up the minimum of these two parameters was the same, but the maximum was increased to 14~cm.

The \(^{252}\)Cf source was used here again, this time distributed uniformly along a line from -10 cm to 10 cm parallel to the face of the detector, i.e. extending beyond the reflecting geometry. The source was biased to preferentially produce neutrons in
the direction of the detector. The source to detector distance for the current optimisation was fixed at 100~cm and
positioned 50~cm above ground level, relative to the \(\sim\)1.2~m high detector.

\subsection{Results}
The measure used to gauge the optimal detector response for this optimisation was the raw \(^3\)He\((n,p)\)\(^3\)H reaction rate
(RR) tallied in the detector volume by MCNP. The MCNP track length estimate of flux (F4 tally) was multiplied by the total \((n,p)\) cross section  of \(^3\)He, thus resulting in a RR in arbitrary units. This response was scored in the detector volume
for each parameter combination. Almost 600 simulations were conducted in total, each running \(10^6\)
particle histories, which resulted in better than 1\% statistical uncertainty in almost every case. Any test cases with
parameter combinations which resulted in overlapping geometry between the air cavity and the
polythene shielding, which resulted in either high statistical errors (\(>20\)\%) or fatal errors in the MCNP simulation, were rejected and given a RR of zero in the following graphical figures and discussion.

\begin{figure}
\center \includegraphics[width=0.9\textwidth]{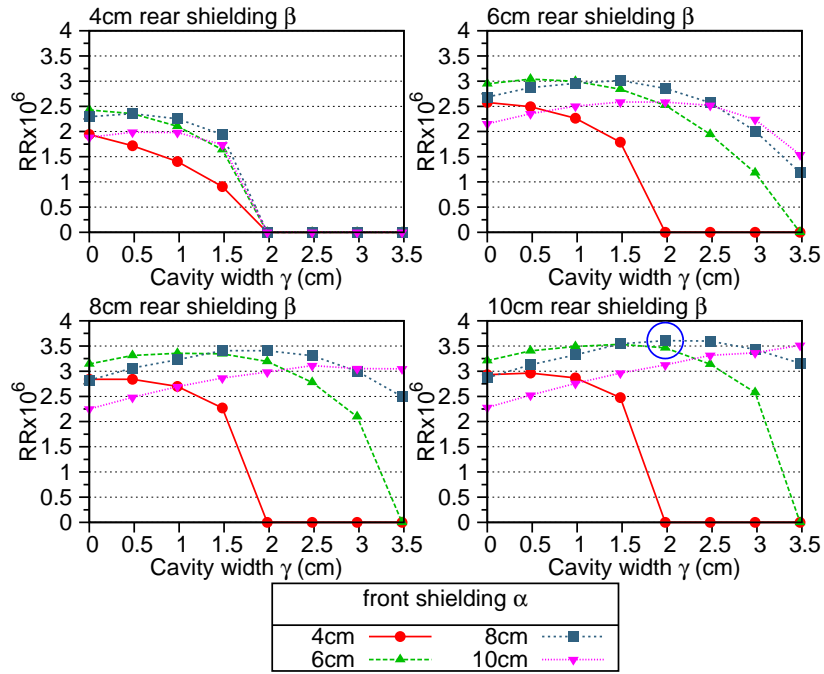}
\caption{\label{cylindrical_results} Reaction rates measured in the helium tube detector as a function of the three design optimisation parameters for a detector surrounded by a cylindrical air cavity. The optimum result is highlighted with a blue circle.}
\end{figure}
\begin{figure}
\center \includegraphics[width=0.9\textwidth]{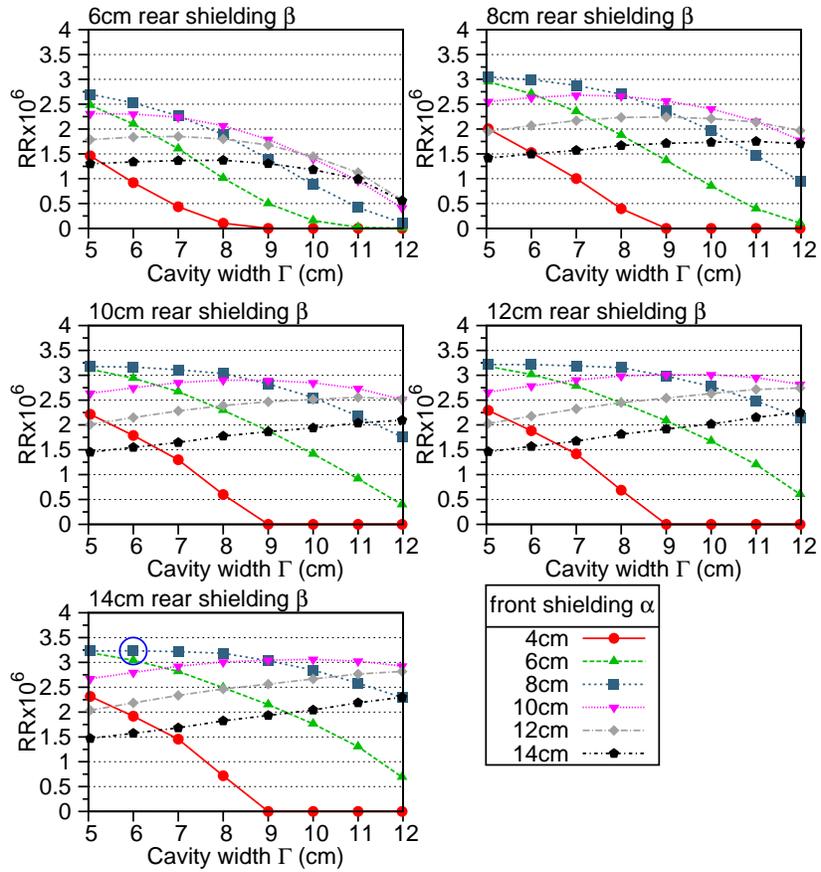}
\caption{\label{linear_results}Reaction rates measured in the helium tube detector as a function of the three design optimisation parameters for a detector surrounded by a rectangular cuboid air cavity. The optimum result is highlighted with a blue circle.}
\end{figure}

Fig.~\ref{cylindrical_results} shows the important results for the cylindrical air gap set-up -- the thin shielding results are omitted because these are either rejects (zero RR), or low RR compared to the thicker shield options. The figure shows in four separate plots the results for rear shielding depths \(\beta\) of 4,6,8, and 10~cm, and in each plot there is a curve of RR versus cavity width \(\gamma\) for these same four thicknesses of front shielding \(\alpha\). From the results it is readily observable that the optimal cylindrical
shell cavity, that with the highest RR, has an air gap of around 2 cm with 8 cm polythene front shielding and 10~cm polythene rear shielding. This best RR from the study is highlighted with a blue circle in the figure, and corresponds to a RR of \(3.61\times10^{-6}\) (rel. err. 0.6\%). For reference, the baseline model used in the optimisation study discussed earlier (section~\ref{optimisation_study}), over which improvement is sought, having no air cavity and uniform 6 cm by 6 cm polyethylene shielding, resulted in a RR of \(2.95\times10^{-6}\) (rel. err.
0.7\%). Thus the inclusion of a cylindrical shell air cavity and additional shielding resulted in an increase in
reaction rate by a factor of 1.22.

In Fig.~\ref{cylindrical_results}, observe also that when the cavity width \(\gamma\) is zero, then, regardless of the thickness of rear-shielding \(\beta\), the optimum \(\alpha\) thickness is not 8~cm, but rather 6~cm. This corresponds to a total polythene thickness of around 3.5~cm between the edge of the steel tank containing the helium and the front face of the detector. This is in good agreement with the result from the main optimisation study (see the discussion in section~\ref{optimisation_results}), where a 4~cm optimum was marginally better than 3~cm (corresponding to \(\kappa\) sizes of 2 and 1~cm, respectively). Now with a 2~cm air gap around the tubes, the optimum \(\alpha\) has increased to 8~cm, maintaining the same 3.5~cm total forward polythene depth. why...?

For the rectangular cuboid cavity simulations also showed an increase in RR over the baseline model, but not as
significant as that for the cylindrical shell case. The sequence of plots in Fig.~\ref{linear_results} shows a subset of the results, where again the thinnest shield configuration options have been omitted.  The highlighted optimal geometry in this case has a 6~cm air cavity, 8~cm polythene
front shielding and 14 cm polythene rear shielding, resulting in a RR of \(3.23\times10^{-6}\) (rel. err. of
0.7\%) -- still a factor of 1.1 over the baseline case.

It is interesting to note that the cross sectional area of the air gap is very similar between the two geometries at 43.7~cm\(^2\) and 52.1~cm\(^2\), respectively, for the cylindrical and rectangular configurations (compare this with a maximum cross sectional area of 124~cm\(^2\) in the rectangular case). This suggests that there is an optimum volume of air to have around a helium tube of a given size. Similarly, identical optimal front polythene shield depths \(\alpha\) of 8~cm are observed, although the shape and thickness of this shielding differs between the models. As noted in the main optimisation study, this forward shielding will strongly influence, through moderation and absorption, the number and energy of the neutrons reaching the detector, so it is not surprising that the values are the same.

Notice also, that for both air gap configurations, the optimum response is obtained with the maximum (of the parameter range considered) rear-shielding thickness. In principle, the response could be improved even more by increasing the rear-shielding further, thereby increasing neutron back-scattering, but at some point there will be a trade-off between marginal response improvement and the cost of more shielding. We return to this point later, during optimisation of the rear-shielding for the full detector system (section~\ref{optimisation_revisited}).

\begin{figure}
\includegraphics[width=1.0\textwidth]{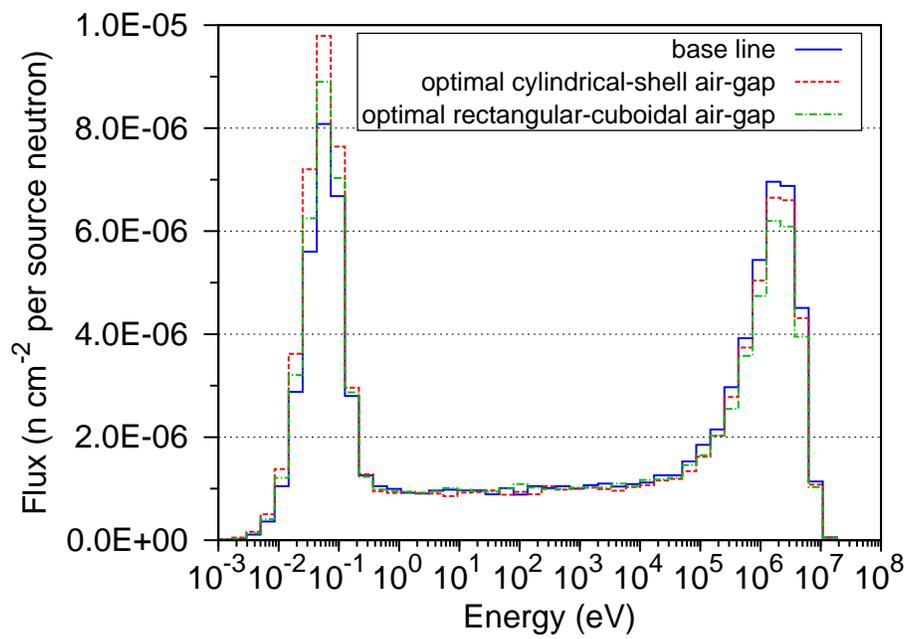}
\caption{\label{he_spectra}Flux spectra recorded in the detector volume for the baseline and optimal detector
configurations.
}
\end{figure}

The spectra recorded in the detector volume, for each of the two optimal models, together with that of the baseline model, are shown in Fig.~\ref{he_spectra}. All three spectra show similar shape profiles, although the thermal flux peak at around 0.07~eV is higher than the baseline for the rectangular optimal case, and higher still with an optimal cylindrical air cavity. The \(^3\)He detector has a \(1/v\) response  and so it is precisely this higher thermal component in the optimal spectra that produces the greater RR compared to the baseline. It is clear from the results that the optimising finds the right balance between thermalizing the fission spectra and retaining a high neutron flux to maximise detector response. The optimal cylindrical shell air gap configuration
has the highest thermal energy neutron flux per source neutron and thus, with contribution from the rest of the
spectra, has the highest reaction rate.

As a final point, note that at for the high-energy peak in the spectra shown in Fig.~\ref{he_spectra}, the optimal cylindrical configuration actually has a higher flux than the optimal rectangular case. This suggests that there is still room for improvement because a further lowering of this peak could increase the thermal peak. The use of other moderator materials, particularly inelastic scatters, to improve the spectrum shaping is undoubtedly an approach worth pursuing but is beyond the scope of this work.

\section{RPM design optimisation revisited}\label{optimisation_revisited}

Using the optimised cylindrical air cavity thickness of \(\sim\)2~cm (in fact 1.98~cm for an overall tube radius of 4.5~cm), into which the standard 2.52~cm radius \(^3\)He tubes are positioned, further optimisation of the full detector system has been carried out. As before, we use the `minimum source to alarm' \(\phi_{src}^{min}\) measure to judge the relative effectiveness of different designs. The thicknesses of the base polythene in front and behind the detector are considered as part of this optimisation process (the \(\kappa\) parameter discussed in section~\ref{parametervariables} already accounts for the front shielding).

\begin{figure}
\includegraphics[width=1.0\textwidth]{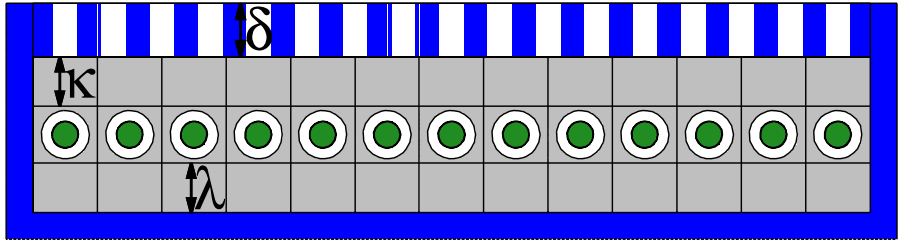}
\caption{\label{geometry_revisited}Top view of detector showing additional optimisation parameter $\lambda $, which allows
for extra polythene shielding behind the helium tubes. Note that in this new improved model the gap between the regions
controlled by $\kappa$ and $\lambda$ and the edge of the air gap surrounding the helium tanks is only 0.02 cm
(previously in section~\ref{optimisation_study} it was 2 cm when there was no air gap).
}
\end{figure}

Fig.~\ref{geometry_revisited} shows the new design for optimisation, including the new variable $\lambda$ that controls the amount of extra polythene shielding between the helium tube units and the start of the rear boronated polythene shielding. As before (section~\ref{parametervariables}), $\kappa$ varies between 0 and 10~cm in the study, while $\lambda$ can be up to 80~cm.

\subsection{Results}
\begin{figure}
\center\includegraphics[width=0.7\textwidth]{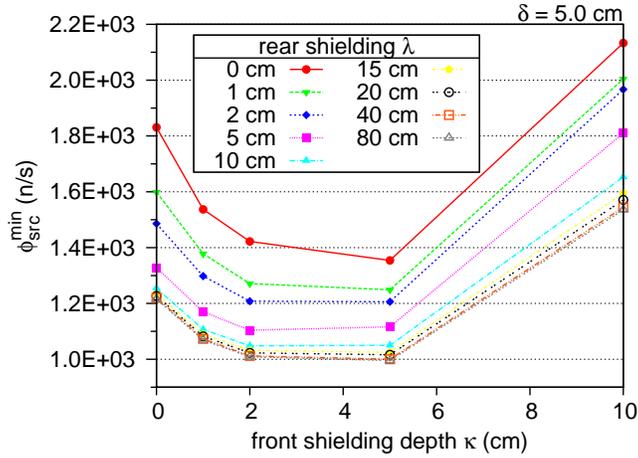}
\caption{\label{revisit_depth}Typical result from the revisited parameter study with a \(\sim\)2~cm air cavity around the helium tubes. The variation in `minimum source to alarm' \(\phi_{src}^{min}\) is shown as a
function of the solid shielding depths ($\kappa$ and $\lambda$) for a fixed collimator depth $\delta$ of 5~cm.
}
\end{figure}

As in the original multi-parameter study of shielding and collimation parameters (section \ref{optimisation_study}), we
considered the optimisation of the shielding thicknesses first, before allowing the other parameters to vary. Fig.~\ref{revisit_depth} shows a typical result from these simulations, in this case for a collimator depth of 5 cm. It is immediately apparent that the best `minimum source alarmed' has fallen from the 1200~n/s observed in the initial parameter study, to around 1000~n/s with the newly optimised helium tube unit cell and air gap -- representing around a 15 \% improvement in optimum performance. In agreement with the detector unit optimisation, the corresponding factor improvement (old/new) is 1.2.

Fig.~\ref{revisit_depth} also demonstrates that, in general, the thicker the rear shielding the better -- a result that is reproduced regardless of the collimator thickness $\delta$. However, above about $\lambda=15$ or 20~cm, there is only marginal improvement in detector performance, and so, as was noted for in the air gap optimisation (section~\ref{baseoptimisation}), there is likely to be a trade-off with sensitivity versus cost of more polythene. In the remainder of the revisited study we take $\lambda=40$~cm as a mid-range thickness that gives almost optimum sensitivity.

In contrast to the previous parameter study, where $\kappa=2$ was optimum, for the new geometry we find that it is an extra shielding thickness $\kappa$ of 5~cm that produces the best detector performance. This reflects the fact that previously, where there was no air gap, there was already 2~cm of polythene between the helium tanks and the collimator before adding extra shielding through $\kappa$. Now, with the 4.5~cm radius tube of air surrounding the tanks, the default shielding is virtually zero. Thus, it is not surprising that $\kappa=5$ becomes optimum, and it is possible that $\kappa=4$ might be slightly better still, although the 1000~n/s best value of \(\phi_{src}^{min}\) is unlikely to change.

\begin{figure}
\center\includegraphics[width=0.7\textwidth]{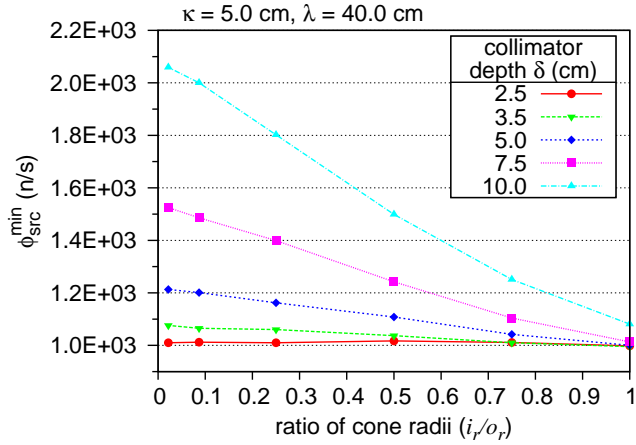}
\caption{\label{revisit_conical}Typical result from the revisited parameter study with a \(\sim\)2~cm air cavity around the helium tubes. The variation in `minimum source to alarm' \(\phi_{src}^{min}\) is shown for fixed values of $\kappa$ and $\lambda$ (5 and 40~cm) as a
function of the collimator profile for various values of collimator depth $\delta$.
}
\end{figure}

Using these optimised values of $\kappa=5$~cm and $\lambda=40$~cm, produces the simulation results shown in Fig.~\ref{revisit_conical} for a new parameter study of the collimator profile. As before, there is no real benefit to making the holes more conical (ratio less than one in the figure), even when the collimator thickness is small. Once again, $\delta=5$~cm produces the best detector sensitivity, in agreement with the original parameter study.

\begin{figure}
\center\includegraphics[width=0.7\textwidth]{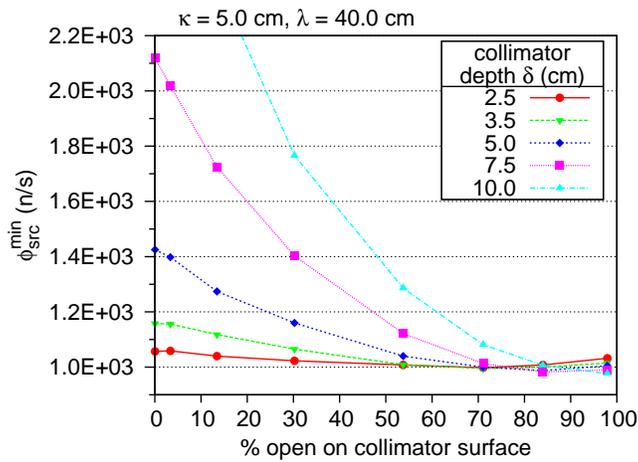}
\caption{\label{revisit_pack}Typical result from the revisited parameter study with a \(\sim\)2~cm air cavity around the helium tubes. The variation in `minimum source to alarm' \(\phi_{src}^{min}\) is shown for fixed values of $\kappa$ and $\lambda$ (5 and 40~cm) as a
function of the \% of the collimator surface occupied by cylindrical holes for  various values of collimator depth $\delta$.
}
\end{figure}

Finally, we vary the hole packing-density, producing the results in Fig.~\ref{revisit_pack}, for the optimised values of $\kappa$ and $\lambda$. The best minimum remains around 1000 n/s, but there is a slight benefit for some of the thicker collimators to having wider holes (greater \% open in the plot).

\section{Summary and Discussion}
This paper describes a sequence of parameters design studies to optimise the performance of a \(^3\)He-based radiation portal monitor (RPM). The sequence of studies, from a preliminary analysis of the impact of the ground material under the RPM, through to detailed optimisations of the various shielding, collimator, and detector-design options,
have built upon one-another and revealed via computational modelling that simple, cost-effective alterations to RPM
design could produce significant improvements in sensitivity to threats present in a background neutron field.

An optimisation study of basic shielding and collimator options demonstrated a
significantly improvement in the response of a generic RPM design to a moving `threat' source in a realistic neutron background field. The parameter study revealed that 2~cm of extra polythene in front of the detector (with a 5~cm thick rear shield) improved performance, which, when combined with a 5~cm thick collimator with cylindrical holes
occupying around 70\% of the exposed surface, could
produce more than a factor of two improved sensitivity compared to an original, unmodified RPM. The optimised,
modified RPM showed \(\sim\)1200~n/s minimum threat detected performance compared to \(\sim\)2600~n/s for the unmodified RPM.

An additional optimisation of the geometry immediately surrounding the \(^3\)He tubes themselves indicated that the inclusion of an air gap, approximately 2~cm thick around the \(^3\)He tubes improved the response (compared to no air) of the detector to the threat source. This knowledge was used to further optimise the shielding and collimation design, and resulted in a further 15\% reduction in the `minimum source alarmed' \(\phi_{src}^{min}\) performance measure (compared to the optimised configuration with no air around the \(^3\)He tanks). In particular, including thick shielding of 20~cm or more behind the detector reduced the response to the background while improving back-scattering of neutrons from the threat source. Combined with a 5~cm thick layer of forward shielding and the same 5~cm thick collimator with a exposed surface about 70\% occupied by cylindrical holes, the sensitivity was improved sufficiently to detect a source strength of only 1000 n/s -- a factor of \(\sim\)2.6 improvement over the unmodified `bare' RPM.

During the optimisation studies it became apparent that a fully-fledged fitting program could be more efficient at identifying optimum configuration,
and would allow all parameters to vary at the same time. A computer program, which would minimise the `minimum source alarmed' objective function by varying the parameter set used to construct the MCNP model would greatly benefit these kinds of studies in the future. Such a development would reveal  a more precise `best parameter set' than those obtained here, and could be targeted at a specific RPM design instead of the generic model considered.
Additionally, the current studies were restricted to \(^3\)He detectors and an unshielded \(^{252}\)Cf neutron source
threat. Further studies should investigate the variation in optimum parameters for both different detectors, such as \(^6\)Li-based scintillators or other non-\(^3\)He based technologies (see~\cite{kouzes2015} for a discussion of the options), and different threat types, including
shielded sources and different spectral sources such as \(^{241}\)Am-Be, \(^{241}\)Am-Li, fissile
metals and oxides for example. In addition, sensitivity of optimum configurations to changing background level, including suppression caused by the passage of large vehicles~\cite{burr2007,wahl2007}, exotic shield/collimation materials, and alternative geometries would be of interest to demonstrate robustness of design.

Finally, the results presented here need to be complemented by experiment, both to confirm the adequacy of the modelling and to prototype the proposed RPM improvements. However, this does not detract from the obvious potential of the modelling approach, which can offer both insight into the factors controlling the efficiency of a RPMs, and suggest practical design changes to improve performance.

\section{Acknowledgements}
This paper describes work that was performed within the CLASP scheme grant entitled `Optimising the neutron
environment of Radiation Portal Monitors (RPM)', which is a collaboration between the United Kingdom Atomic Energy Authority, Glasgow University (Val O'Shea), and Sheffield University (John McMillan). The United Kingdom Atomic Energy Authority was awarded funding through the United Kingdom Science and Technology Facilities Council STFC (grant no. STK000152/1) in 2012. The authors would like to thank Dick Lacey of the Home Office Scientific Development Branch and Antonin Vacheret of Oxford University for useful discussions in support of this work.
To obtain further information on the data and models underlying this paper please contact PublicationsManager@ccfe.ac.uk

\bibliographystyle{elsarticle-num}
\bibliography{clasp_paper}

\end{document}